\documentclass[review]{article}
\usepackage{lineno,hyperref}
\modulolinenumbers[5]
\usepackage{amsmath}

\usepackage{amssymb}
\usepackage{xcolor}
\usepackage{bbm}
\usepackage{amsthm}
\usepackage{graphicx}
\usepackage{subcaption}
\usepackage{multirow}%
\usepackage{amsmath,amsfonts}%
\usepackage{mathrsfs}%
\usepackage{xcolor}%
\usepackage{textcomp}%
\usepackage{manyfoot}%
\usepackage{booktabs}%
\usepackage{algorithm}%
\usepackage{algorithmicx}%
\usepackage{algpseudocode}%
\usepackage{listings}%
\usepackage{bbm}
\usepackage{subcaption}
\usepackage{geometry}

\newtheorem{theorem}{Theorem}
\newtheorem{lemma}{Lemma}
\begin{document}

\title{Estimating parameters of continuous-time multi-chain hidden Markov models for infectious diseases\thanks{This project was supported by the LabEx NUMEV (ANR 2011-LABX-076) within the I-SITE MUSE and the French Agence Nationale de la Recherche (ANR), under grant ANR-21-CE40-005 (project HSMM-INCA)}}

\author{Ibrahim Bouzalmat\footnote{IMAG, Univ Montpellier, CNRS, Montpellier, France} \and Benoîte de Saporta\footnote{IMAG, Univ Montpellier, CNRS, Montpellier, France} \and Solym M. Manou-Abi\footnote{IMAG, Univ Montpellier, CNRS, Univ Mayotte, Montpellier, France}}

\date{}

\maketitle

\begin{abstract}
This study aims to estimate the parameters of a stochastic exposed-infected epidemiological model for the transmission dynamics of notifiable infectious diseases, based on observations related to isolated cases counts only. We use the setting of  hidden multi-chain Markov models and adapt the Baum-Welch algorithm to the special structure of the multi-chain. From the estimated transition matrix, we retrieve the parameters of interest (contamination rates, incubation rate, and isolation rate) from analytical expressions of the moments and Monte Carlo simulations. The performance of this approach is investigated on synthetic data, together with an analysis of the impact of using a model with one less compartment to fit the data in order to help for model selection.
\end{abstract}

\section{Introduction}
Infectious diseases, characterized by varying degrees of person-to-person and environmental contamination rates, incubation periods, and recovery rates, pose significant challenges to public health. These diseases, often endemic in nature, are particularly prevalent in contexts characterized by poverty, limited access to potable water, and inadequate sanitation conditions \cite{gasem2001poor, hosoglu2006risk, king1989community}. Examples of such diseases include, Typhoid fever, Severe Acute Respiratory Syndrome (SARS), Acute Gastroenteritis, and common childhood diseases such as adenovirus, chickenpox, and influenza \cite{gerba2009environmentally,monack2012salmonella,virlogeux2015incubation,wikswo2015outbreaks}.

In the field of infectious disease research, inferring and estimating the parameters of processes governing transmission dynamics is of paramount importance. Numerous previous studies have contributed to the development of methodologies for parameter estimation and inference in various disease contexts. Early works by Anderson and May \cite{anderson1991infectious} pioneered the use of mathematical models and likelihood-based methods to estimate key parameters governing disease transmission. Since then, researchers have made substantial progress in refining these methodologies and adapting them to specific diseases. Furthermore, recent advancements in statistical inference techniques have led to the development of more sophisticated estimation methods \cite{held2019handbook,held2005statistical,noufaily2013improved,reiner2013systematic}. For instance, \cite{gilks1995markov} demonstrated the effectiveness of Markov chain Monte Carlo (MCMC) algorithms in exploring complex parameter spaces and improving estimation precision. These methods have become widely adopted in the field of infectious disease modeling. In recent years, there has been a growing emphasis on the integration of multiple data sources for parameter estimation. In \cite{lekone2006statistical}, the authors proposed a framework that combines information from multiple surveillance systems, incorporating both case counts and prevalence data, to enhance the accuracy of parameter estimation. 

In this study, we present a parametric framework designed to estimate the parameters that characterize the dynamics of the spread of infectious diseases, in the special case where the number of infected remains very low compared to the susceptible population, so that stochastic models are relevant. This framework includes a two-compartment model, which allows us to track the count of individuals exposed and infected by the disease. Specifically, exposed individuals are characterized as those who have encountered the disease-causing bacterium or virus but have yet to exhibit overt symptoms. During this latent period, individuals are typically considered non-contagious. In contrast, infected individuals are those who have progressed beyond the incubation period and are now symptomatic, actively carrying and transmitting the disease \cite{gauld2018typhoid,olsen2003outbreaks}. Susceptible are not counted as we are interested in modeling low incidence diseases. Isolated or recovered individuals are partially observed and added to the model when necessary. 
The stochastic compartmental model described by a pure jump process is well-suited for capturing the discrete and irregular nature of disease transmission events with few interpretable parameters. The objective of our study is to estimate the parameters of this exposed-infected model, namely person-to-person and exogenous contamination rates, incubation rate, and isolation rate. However, our approach faces specific challenges, making the estimation problem complex due to two major factors. Firstly, the observations are not continuous over time, but aggregated over fixed periods of time (typically one day or one week). Additionally, for each period the number of exposed or infected individuals is not observed, only cumulative counts of newly reported cases over the period are available. 

Although this type of observation is common in public health settings, few works investigate parameter estimation for stochastic processes with this observation scheme. We have recently addressed this challenging question for a simpler class of models with only one compartment  \cite{bouzalmat2024parameter}. The main difficulty with adding an exposed compartment is that there is no analytical form for the transition matrices of the skeleton chain taken at discrete regular intervals. Another interesting recent work is \cite{CL22} where the author use Hawkes processes instead of compartmental models to model disease outbreaks with similar characteristics.

In the present study, we build on  \cite{bouzalmat2024parameter} and extend the methodology used to include both exposed and infected compartments, thus incorporating an additional parameter to describe the incubation period of the epidemic. By introducing this parameter and accounting for individuals in the exposed compartment, we capture a more comprehensive representation of the disease's life cycle and transmission dynamics. In our estimation method, the first step involves studying the moments of the two-dimensional exposed-infected process. The objective is to derive analytical expressions representing their expected values in relation to the model parameters. Importantly, this step is performed under specific stability conditions, which are essential to ensure the validity of our methodology. The second pivotal step takes into account the fact that only new isolated cases are recorded, and we consider a framework of hidden multi-chain Markov model (HMCMM) or coupled hidden Markov model, which is considered an extension of the classical hidden Markov model. This extension allows capturing complex dependencies between different sequences of states, making them particularly useful in many applications \cite{brand1997coupled,ghosh2017septic,kumar2017coupled, sarno2017coupled}. Next, we adapt the Baum-Welch algorithm \cite{baum1966statistical,turin1998unidirectional,yang2017statistical}, which enables the estimation of the transition probability of the coupled hidden Markov chain from our non-standard observation scheme. Adaptation is necessary as the hidden chain has a special structure. We then estimate the moments of the process from the estimated transition matrix using Monte Carlo simulations \cite{harrison2010introduction,mooney1997monte} and use the analytical formulas obtained in the first step to obtain parameter estimates. The global performance of the estimation is evaluated through simulations both in the case where data come from a two-compartment model and in the case where data come from a single compartment model, to study the impact of model choice. 

This paper is structured as follows. In Section \ref{sec2}, we set the stochastic exposed-infected model and present our special observation scheme. Section \ref{sec3} specifies our parameter estimation procedure. In Section \ref{sec4}, we leverage numerical simulations to both evaluate the achieved results and conduct a model choice analysis. Finally, the article concludes \ref{sec5} with a brief summary of the main findings and potential directions for future research. The proofs and additional technical results are gathered in the Appendix. The codes developed for this study are available at \url{https://plmlab.math.cnrs.fr/bouzalma/mchmm.git}.
%
\section{Exposed-Infected model and observation scheme}
\label{sec2}
%
In this section, we begin describe the stochastic model employed for counts of both exposed and infected individuals in Section \ref{sec:model}. Following that, we specify the observation process in Section \ref{sec:obs}.
%
\subsection{Mathematical formulation of the model}
\label{sec:model}
%
In this study, we focus on a Susceptible-Exposed-Infectious-Isolated model for notifiable diseases with low prevalence. Thus, we only consider the exposed and infectious compartments, considering that changes in the number of susceptible and isolated individuals are negligible and do not influence the disease propagation. For similar reasons, deaths are not counted either.

\begin{figure}[t]
\begin{center}
\includegraphics[scale=0.12]{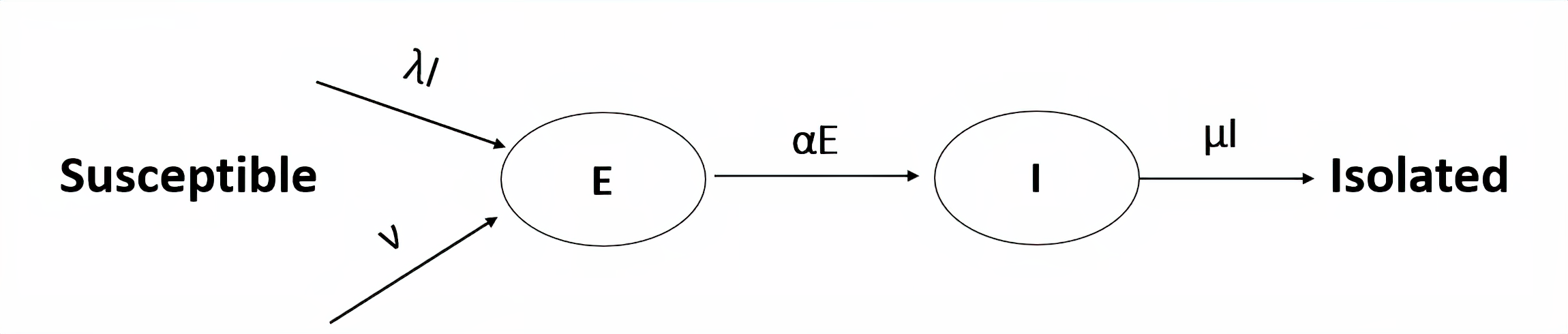}
\caption{\textbf{Exposed-Infected two-compartment model and key parameters.}}
\label{Fig2.1}
\end{center}
\end{figure}

As mentioned in the introduction, the disease dynamics are characterized by an incubation phase, a contagious phase, and an isolation phase. The incubation phase, corresponding to compartment $E$ for exposed, starts when the individual has come into contact with the disease but is not yet infectious. Contact with the disease may come from two different sources: either from contact with an infected individual with rate $\lambda$ or form the environment or some other exogenous source with rate $\nu$. An exposed individual becomes infected, i.e. moves to compartment $I$, with incubation rate $\alpha$. An infected individual exhibits symptoms of the disease and may contaminate susceptible individuals. With isolation rate $\mu$, the infected individual goes to a medical center and is diagnosed with the disease. They are then reported to the authorities and isolated to be treated, therefore no more contaminating, see Fig.\ref{Fig2.1}.
For instance, for typhoid fever, isolation corresponds to hospitalization with antibiotic treatment and recovery typically occurs within 1 to 5 days, depending on the severity of the infection and the response of the patient to treatment  \cite{frenck2004short,wain2001quantitation}.
For simplicity, we consider that declaration and isolation are simultaneous, and that all cases are reported.

We propose a two-compartment stochastic model to describe this transmission dynamics.
Let  $ (E_t,I_t )_{t \geq 0 }$ be a continuous-time bivariate Markov process taking values in $\mathbb{S} =\mathbb{N}\times\mathbb{N}$, where $E_t$ and $I_t$ respectively represent the number of exposed and infected individuals at time $t$ within the population.
Given the current state $(e,i)$, the possible transitions of our process are: 
\begin{itemize}
\item $(e,i)\rightarrow (e+1,i)$ with a rate of $\lambda i + \nu$: a new individual is exposed,
\item $(e,i)\rightarrow (e-1,i+1)$ with a rate of $\alpha e $: an individual moves from the exposed compartment to the infected one,
\item $(e,i)\rightarrow (e,i-1)$ with a rate of $\mu i$: an infected individual is isolated.
\end{itemize}
Thus, the transition rates matrix $Q$ of the process is given by
\begin{align}\nonumber\label{eq4.1}
Q_{(e,i),(e+1,i)}&= \lambda i + \nu, \\ \nonumber
Q_{(e,i),(e-1,i+1)}&= \alpha e \mathbf{1}_{e>0}, \\   
Q_{(e,i),(e,i-1)}&=\mu i\mathbf{1}_{i>0},\\\nonumber
Q_{(e,i),(e,i)}&= - (( \lambda + \mu ) i + \alpha e + \nu ),
\end{align}
and $Q_{(e,i),(e',i')}=0$ in all other cases. 

For any $(e,i)$ and $(e',i')$ in $\mathbb{S}$ and $t\geq 0$, benote by
$$
p_{(e,i)(e',i')}(t)=\mathbb{P}\big((E_t,I_t )= (e',i') \mid (E_0,I_0 ) = (e,i) \big)
 \quad (e',i') \in \mathbb{S}, t \geq 0,
$$
the transition probability of the process $(E_t,I_t )_{t \geq 0 }$. They satisfy the forward Kolmogorov equation
\begin{align}\label{eq2.1}
\lefteqn{\dfrac{\mathrm{d} }{\mathrm{d} t}p_{(e,i),(e',i')}(t)}\nonumber\\
&= \left(\lambda i' + \nu \right) p_{(e, i),(e'-1,i')}(t)+\alpha (e'+1) p_{(e,i),(e'+1,i'-1)}(t) \\ \nonumber
& \quad+ \mu (i'+1) p_{(e, i),(e',i'+1)}(t)-\left(( \lambda + \mu )i'+\alpha e'  +\nu\right) p_{(e, i),(e',i')}(t), 
\end{align}
with the initial condition $p_{(e,i),(e,i)}(0)=1 $ and $p_{(e,i),(e',i')}(0)=0 $ for $(e',i')\neq (e,i)$. 
%
\subsection{Observation scheme}
\label{sec:obs}
Let $Y_{(s,t]}$ be the cumulative count of newly isolated cases within the time interval $(s,t]$. In other words, $Y_{(s,t]}$ corresponds to the number of jumps of amplitude $-1$ of $I$ over the time interval $(s,t]$. Set  $N_0=0$ and $N_t=Y_{(0,t]}$ for $t>0$. Then $N_t$ is the total number of isolated since the beginning. We use the notation $Y_{(0,t]}$ to emphasize that the count starts at $0$. The only observation available are the cumulated counts of isolated over fixed periods of time, denoted by $(Y_n : = Y_{((n-1) \Delta t, n\Delta t]}=N_{n\Delta}-N_{(n-1) \Delta t})_{n\in\mathbb{N}^*}$, where $\Delta t$ is a fixed time step (typically $1$ day or $1$ week). 

The joint process $(E_t,I_t, Y_{(0,t]})_{t\geq 0}$ is still a Markov process and is characterized by the following transition probabilities. For $t>0$, set
\begin{align*}
p_{(e,i),(e',i',y)}(t):= \mathbb{P} \left((E_t,I_t) =(e',i'), Y_{(0,t]} = y | (E_0, I_0)=(e,i)\right).
\end{align*} 
These probabilities satisfy a Kolmogorov equation.
\begin{lemma}\label{lemma2.1}
For $t>0$ and $e,i,e',i',y\in\mathbb{N}$, we have 
\begin{align}\label{eq2.2}
\lefteqn{\dfrac{d }{dt}p_{(e,i),(e',i',y)}(t) }\nonumber\\
&= (\lambda i + \nu )  p_{(e+1,i),(e',i',y)}(t) + \alpha e   p_{(e-1,i+1),(e',i',y)}(t)\\ \nonumber
&\quad + \mu i  p_{(e,i-1),(e',i',y-1)}(t) - ((\lambda  + \mu ) i + \alpha e + \nu )p_{(e,i),(e',i',y)}(t),
\end{align} 
for $i\leq i'+y$, with $p_{(e,i),(e',i',y)} = 0$ when $i > i'+y$ and with the initial condition $p_{(e,i),(e',i',y)}(0) = \mathbf{1}_{i=i'} \mathbf{1}_{y=0}$.
\end{lemma}
The proof of this lemma is provided in Appendix \ref{proof::Lemma1}.
Equations \eqref{eq2.1} and \eqref{eq2.2} do not have a closed-form solution, making it impossible to express the marginal distribution of the observations in terms of the parameters $\lambda, \mu, \alpha$, and $\nu$, or to obtain explicit expressions for the transition probabilities using moment-generating functions as done in \cite{bouzalmat2024parameter}. As a result, direct maximum likelihood estimation is not feasible. Although the probability-generating function of the process $(E_t,I_t)_{t\geq 0}$ can be calculated,
it is not possible to identify the distribution of $(E_t,I_t)_{t\geq 0}$. To overcome these challenges, we will adopt an alternative estimation method using our specific observation sequence $(Y_n)$. 
%
\section{Estimation} 
\label{sec3}
%
In this section, we present our method for estimating the parameters of the exposed-infected model. To begin, in Section \ref{sec:strategy} we explain our estimation strategy and the steps involved in this method. Then we detail the main steps: derivation of moment estimators in Section \ref{sec:moment}, inference of the transitions matrix for the hidden multi-chain model associated with the sequences of exposed, infected, and isolated states and finally estimators of the parameters of interest in Section \ref{subsect::3.4}. 
%
\subsection{Estimation strategy}
\label{sec:strategy}
%
Our estimation problem belongs to the class of hidden information problems. The first step in our estimation method is to consider the case where the variables $E_t$, $I_t$ and $Y_{(0,t]}$ are observed.
We study the average behavior of the joint process $(E_t,I_t,Y_{(0,t]})_{t \geq 0}$, and are able to express their moments explicitly as a function of the parameters $\lambda, \mu, \alpha$ and $\nu$. Under some stability condition, 
the limit as $t$ goes to infinity of the system of expectations is invertible, which allows us to deduce an explicit expression for the parameters in terms of the limits of the moments of the process $(E_t,I_t,Y_{(0,t]})_{t \geq 0}$ when $t$ goes to infinity. 

The second phase takes into account the missing data framework by considering the sequences of exposed-infected-isolated states as a hidden multi-chain Markov model. We rewrite the characteristics of the model taking into account our the particular structure of the multi chain and adapt Baum Welsh's algorithm for estimating the transition probability of the coupled hidden Markov chain from the sequence of observations. 

We then estimate the moments of the process using Monte Carlo simulations from the estimated transition matrix and use the analytical formulas obtained in the first step to obtain estimates of the original parameters through a \emph{plug-in} approach.
%
\subsection{Moment estimators}
\label{sec:moment}
%
We use the method of moments to estimate the parameters  $(\lambda, \mu, \alpha,\nu)$ when the whole process $(E_t,I_t,Y_{(0,t]} )_{t \geq 0 }$ is observed for some large date $t$.
\begin{theorem}\label{th3}
If $\lambda < \mu$, then the parameters $\lambda,\mu,\alpha$ and $\nu$ are given by the following formulae 
\begin{eqnarray}\nonumber  \label{eq3.2}
\lambda &=& \dfrac{ N^{\star}\left(\frac{R^*}{E^* I^*} -1 \right) \left( E^* + I^*\right)}{I^* \left( 1+ \left(\frac{R^*}{E^* I^*} -1 \right) \left( E^* + I^*\right) \right)}, \\ 
 \mu &=& \dfrac{N^{\star}}{I^*}, \\ \nonumber
  \alpha &=& \dfrac{N^{\star}
  }{E^*}, \\ \nonumber
 \nu &=& I^* \left( \mu- \lambda \right),
\end{eqnarray}
where $$N^{\star}  = \lim_{t \rightarrow +\infty} \dfrac{\mathbb{E}_{(e_0,i_0)} [ Y_{(0,t]}]}{t}, \ \ 
  E^{\ast} = \lim_{t \rightarrow +\infty} \mathbb{E}_{(e_0,i_0)} [ E_t],
$$ 
$$    I^* = \lim_{t \rightarrow +\infty} \mathbb{E}_{(e_0,i_0)} [ I_t],  \ \ \text{ and }  \   R^* = \lim_{t \rightarrow +\infty} \mathbb{E}_{(e_0,i_0)} [ E_t I_t], $$
  for $e_0,i_0,y_0 \in \mathbb{N}$.
\end{theorem}
The proof is given in Appendix \ref{proof::Th3.1}.
The consistency and asymptotic normality of the estimators for $(\lambda, \mu, \alpha, \nu)$ are thus assured as long as the estimators for $E^*, I^*, R^*$, and $N^{\star}$ are consistent and asymptotically normal.
%
\subsection{Hidden Markov Multi-chain model}
\label{subsect::3.4}
%
In our observation framework, the values of $(E_t,I_t)$ are hidden at all times, and only the process $Y_{(0,t]}$ is observed at dates $n\Delta t$ as $Y_{(0,N\delta t]}=\sum_{n=1}^{N}Y_n$. We consider discrete-time Markov chain $(E_n,I_n)$ where, with a slight abuse of notation, $E_n=E_{n \Delta t}$ and $I_n=I_{n \Delta t}$. Unfortunately, it is not possible to obtain analytical expressions for the transition probability matrix of this chain as functions of our parameters. Therefore, following the idea introduced in \cite{bouzalmat2024parameter}, we make use of the Hidden Markov Model (HMM) framework to obtain estimations of the transition matrix from the available observations $Y_n$. Note that the standard algorithm must be adapted to the special structure of our chain.

There are to main difficulties to be taken into account. First, the hidden chain is two-dimensional and exposed and infected are coupled, so we are in the framework of hidden multi-chain Markov models (HMCMM). Second, the observation $Y_n$ at time $n$ depends on both the infected count $I_n$ at time $n$ and $I_{n-1}$ at time $n-1$. Indeed, it depends on the whole path of $I_t$ for $n< t\leq n+1$ which is a non standard emission pattern.

We can reduce our HMCMM to an HMM with a standard observation scheme by introducing the three-dimensional chain $(X_n)_{n\in\mathbb{N}^*} = (E_{n-1},I_{n-1},I_n)_{n\in\mathbb{N}^*}$ with values in the state space $\mathbb{N}^3$. However this chain has a special structure as the last two components are the same chain at consecutive time steps. Ignoring this structure yields identifiability problems and irrelevant estimates. The special structure of the HMM leads to the following specific form for its parameters.
\begin{lemma}\label{lemme2}
 The process $(X_n,Y_n)_{n\in\mathbb{N}^*}$ is a hidden Markov model whose characteristics are given by the triple $M = (Q, \psi, \rho)$, where  
\begin{enumerate}
\item the transition probability matrix $Q$ of the hidden chain $(X_n)$ is 
 $$ Q_{(e,i,j),(e',i',j')}= \dfrac{p_{(e',i'),(\centerdot,j')}p_{(e,i),(e',i')} }{p_{(e, i),(\centerdot,j)}}\delta_{i'=j},$$
for $(e,i,j) , (e',i',j') \in \mathbb{N}^3$, where the $ p_{(e,i),(e',i')} = p_{(e,i),(e',i')}( \Delta t )$ are the transition probabilities of  $(E_{n},I_{n})_{n\in\mathbb{N}}$ and $p_{(e, i),(\centerdot,j)} = \sum_{k}p_{(e, i),(k,j)}$;
\item the emission probability of the process $Y$ given the process $X$ is 
 $$\psi_{(e,i,j)}(y)= \mathbb{P}\left(Y_n = y | X_n=(e,i,j)\right)= \dfrac{p_{(e,i)(\centerdot,j,y)}}{p_{(e,i),(\centerdot,j)}},$$
for $e,i,j,y \in \mathbb{N}$,  where $ p_{(e,i)(\centerdot,j,y)}= \sum_{k}p_{(e,i)(k,j,y)} (\Delta t)$;
\item the initial distribution $\rho$ of the  state process $X$ is 
$$\rho_{e,i,j} =\mathbb{P}(X_1 = (e,i,j) )=  p_{(e, i),(\centerdot,j)} \pi_{(e,i)},$$
for $e,i,j \in \mathbb{N}$, where $ \pi$ is the initial distribution of the chain $(E_{n},I_{n})_{n\in\mathbb{N}}$.
\end{enumerate}
\end{lemma}
The proof is straightforward and left to the reader.
In the following, we aim to estimate the parameters of the hidden Markov model $(X_n,Y_n)$ and more specifically the transition matrix of the chain $(E_{n},I_{n})$, by maximizing the likelihood function $ \mathbb{P}(Y_{1}= y_{1},\ldots, Y_{T}= y_{T}| M )$ using an adapted version of the Baum-Welch or forward-backward algorithm. We have rewritten the main functions of the algorithm, adapting the recursive formulae to the particular structure of our model in order to obtain estimators of the transition probabilities $p_{(e,i),(e',i')}$.

For a given sequence of observations $(y_1,\ldots,y_T)$, the Forward probability can be calculated by the following recursion. For any $e,i,j \in \mathbb{N}$ and $2 \leq t \leq T$, one has
$$
\begin{cases}
\alpha_{(e,i,j)}(1)= p_{(e,i),(\centerdot,j)} \pi_{(e,i)} \psi_{(e,i,j)}\left(y_{1}\right), \\
\alpha_{(e,i,j)}(t)=\psi_{(e,i,j)}\left(y_{t}\right) \sum_{(e',i')\in \mathbb{N}^2}\dfrac{p_{(e,i),(\centerdot,j)}p_{(e',i'),(e,i)} }{p_{(e', i'),(\centerdot,i)}} \alpha_{(e',i',i)}(t-1).
\end{cases}
$$
Similarly, the Backward probability can be evaluated by the recursive formula
$$
\begin{cases}
\beta_{(e,i,j)}(T)= 1, \\
\beta_{(e,i,j)}(t)=  \sum_{(e',j')\in \mathbb{N}^2}\dfrac{p_{(e',j),(\centerdot,j')}p_{(e,i),(e',j)} }{p_{(e, i),(\centerdot,j)}} \psi_{(e',j,j')}\left(y_{t+1}\right) \beta_{(e',j,j')}(t+1),
\end{cases}
$$
for $e,i,j \in \mathbb{N}$ and $1 \leq t \leq T-1$.
Using the Forward and Backward probabilities, the likelihood of the observations given model $M$ is
$$ \mathbb{P}(Y_{1}= y_{1},\ldots, Y_{T}= y_{T}| M ) = \sum_{(e,i,j)\in \mathbb{N}^3} \alpha_{(e,i,j)}(t) \beta_{(e,i,j)}(t). $$
The following result presents the iterative scheme of the adapted Baum-Welch algorithm.
\begin{theorem}\label{th4}
Given the model $M^n = (Q^n,\psi^n,\rho^n)$, the maximum likelihood estimates of $M^{n+1} = (Q^{n+1},\psi^{n+1},\rho^{n+1})$ given the observations $(y_1,\ldots,y_T)$ are given, for $e, i, j,e',i',j' \in \mathbb{N} $, by
\begin{align*}
Q_{(e,i,j),(e',i',j')}^{n+1} &= \dfrac{\sum_{t=1}^{T-1}\xi^n_{(e,i,j),(e',i',j')}(t)}{\sum_{t=1}^{T} \gamma^n_{(e,i,j)}(t) } \delta_{i'=j}, \\ 
\psi_{(e,i,j)}^{n+1}(o) &= \dfrac{\sum_{t=1}^{T} \mathbf{1}_{o_t = o} \gamma^n_{(e,i,j)}(t)}{\sum_{t=1}^{T} \gamma^n_{(e,i,j)}(t)},\\
\rho_{e,i,j}^{n+1}&= \gamma^n_{(e,i,j)}(1),\\
p_{(e,i),(e',i')}^{n+1}&= \dfrac{\sum_{j'\in\mathbb{N} }Q^{n+1}_{(e,i,j),(e',j,j')}\rho_{e,i,j}^{n+1} }{\sum_{j\in\mathbb{N}}\rho_{e,i,j}^{n+1}},
\end{align*}
where 
\begin{align*}
\lefteqn{\xi_{(e,i,j),(e',i',j')}^n(t)}\\
&= \dfrac{\alpha^n_{(e,i,j)}(t)p^n_{(e',i'),(\centerdot,j')}p^n_{(e,i),(e',i')}\psi^n_{(e',i',j')}(o_{t+1}) \beta^n_{(e',i',j')}(t+1) }{ p^n_{(e, i),(\centerdot,j)} \sum_{(e,i,j)\in \mathbb{N}^3} \alpha_{(e,i,j)}^n(t) \beta_{(e,i,j)}^n(t)}\delta_{i'=j},\\
\gamma_{(e,i,j)}^n(t)&= \dfrac{\alpha_{(e,i,j)}^n(t) \beta_{(e,i,j)}^n(t)}{ \sum_{(e,i,j)\in \mathbb{N}^3} \alpha_{(e,i,j)}^n(t) \beta_{(e,i,j)}^n(t)},
\end{align*}
and $\alpha_{(e,i,j)}^n(t), \beta_{(e,i,j)}^n(t)$ are the Forward and Backward probabilities at iteration $n$ defined above.
\end{theorem}
The proof of this theorem is given in Appendix \ref{proof::th3.2}. At each iteration, the Baum-Welch algorithm updates the model parameters using the formulae given in Theorem \ref{th4} and re-evaluates the log-likelihood of the observations, until convergence is achieved. Obtaining the optimal transition matrix $\hat{p}$ allows the Markov chain $(E_n,I_n)$ to be simulated with a sufficiently large $n$ to retrieve empirical estimations of the moments $E^*,I^*$ and $R^*$. Finally, Theorem \ref{th3} can be used to estimators $\hat{\lambda}^n, \hat{\mu}^n, \hat{\alpha}^n$ and $\hat{\nu}^n$ of the parameters of the exposed-infected model $\lambda, \mu, \alpha$ and $\nu$ from the estimated moments.
%
\section{Performance estimation}
\label{sec4}
%
In this section, we study the performance of the estimators $\hat{\lambda}^n, \hat{\mu}^n, \hat{\alpha}^n$ and $\hat{\nu}^n$ on synthetic data. Estimation errors come from several sources that we investigate separately. 
The first source of error is the replacement of the limit moments $E^*$, $I^*$, $N^\ast$ and $R^*$ by estimations. This is investigated in Section \ref{sec:complete}. The second source of error comes from truncating the state space of the HMM $(X_n,Y_n)_{n\in\mathbb{N}^*}$ in order to have a finite-size transition matrix. This is discussed in Section \ref {sec:trunc}.
In Section \ref{HMM}, we apply the full procedure: truncation of the state space, HMM estimation of the transition matrix, Monte Carlo estimates of the limit moments and retrieval of the original parameters.
Finally, in a Section \ref{sec:choice}, we also investigate model selection between models with and without the exposed compartment.
%
\subsection{Estimation of the limit moments}
\label{sec:complete}
%
We generate $N_{MC}=10000$ samples of the random variables $(E_H, I_H, Y_{(0,H]})$ for fixed parameter values $\lambda=0.05, \mu=0.2, \alpha=0.1$, and $\nu=0.015$ and several values of the truncation horizon $H$: $H=1000$, $H=5000$ and $H=10000$. 

\paragraph{Limit moments of $E$ and $I$} The limit moments $E^*$, $I^*$, and $R^*$ are estimated by the empirical moments $\overline{E_{H}}_{N_{MC}}$, $\overline{I_{H}}_{N_{MC}}$, and $\overline{EI_{H}}_{N_{MC}}$ at time $H$. On the one hand, it is well known that the error coming from Monte Carlo simulations is of the order of magnitude of $N_{MC}^{-1/2}$. On the other hand, the error coming from truncating at time $H$ is also known from Eq. \eqref{eqC.2} in the proof of Theorem \ref{th3} and is of order of magnitude $e^{-cH}$ for some explicit constant $c$ depending on the parameters (and different for each moment). Hence for large enough $H$, the truncation error should be negligible compared to the Monte Carlo error, as illustrated in Table \ref{table1}.
%
\begin{table}[t]
\begin{center}
\caption{Estimates of the limit moments $E^*$, $I^*$ and $R^*$ (with $95\%$ confidence interval). True values are $E^*=0.2$, $I^*=0.1$ and $R^*=0.042$}
\label{table1}
\begin{tabular}{@{}lrrrr@{}}
\hline
&\multicolumn{3}{c}{$N_{MC}=100$}\\
\cline{2-5}
Estimator& $H=1000$ &$H=5000$&  $H=10000$&$H=100000$ \\
\hline
$\overline{E_{H}}_{N_{MC}}$  & $0.16\ (0.082;0.237)$  & $0.25\ (0.156;0.343)$& $0.22\ (0.129;0.310)$&  $0.21\ (0.131;0.289)$\\
$\overline{I_{H}}_{N_{MC}}$  & $0.14\ (0.056;0.223)$ & $0.11\  (0.042;0.177)$ & $0.10\ (0.034;0.165)$&$0.10\ (0.040;0.160)$\\
$\overline{EI_{H}}_{N_{MC}}$  & $0.08\ (\phantom{0.20}0;0.200)$  & $0.04\ (\phantom{0.20}0;0.087)$& $0.05\ (\phantom{0.20}0;0.108)$&$0.04\ (\phantom{0.20}0;0.074)$ \\
\hline 
&\multicolumn{3}{c}{$N_{MC}=1000$}\\
\cline{2-5}
Estimator& $H=1000$ &$H=5000$&  $H=10000$& $H=100000$\\
\hline
$\overline{E_{H}}_{N_{MC}}$  & $0.186\ (0.158;0.213)$  & $0.210\ (0.182;0.237)$& $0.203\ (0.175; 0.229)$&$0.198\ (0.172;0.224)$ \\
$\overline{I_{H}}_{N_{MC}}$  & $0.090\ (0.079;0.119)$ & $0.094\ (0.074;0.113)$ & $0.097\ (0.077;0.116)$&$ 0.101\ (0.082;0.119)$\\
$\overline{EI_{H}}_{N_{MC}}$  & $0.038\ (0.022;0.053)$  & $0.047\ (0.027;0.066)$& $0.045\ (0.026;0.063)$ &$ 0.043\ (0.026;0.057)$\\
\hline 
&\multicolumn{3}{c}{$N_{MC}=10000$}\\
\cline{2-5}
Estimator& $H=1000$ &$H=5000$&  $H=10000$& $H=100000$\\
\hline
$\overline{E_{H}}_{N_{MC}}$  & $0.204\ (0.195;0.213)$  & $0.206\ (0.197;0.214)$& $0.200\ (0.192;0.208)$& $0.200\ (0.194;0.206)$ \\
$\overline{I_{H}}_{N_{MC}}$  & $0.107\ (0.099;0.115)$ & $ 0.100\ (0.093;0.106)$ & $0.100\ (0.093;0.106)$&$ 0.100\ (0.095;0.105)$ \\
$\overline{EI_{H}}_{N_{MC}}$  & $0.041\ (0.035;0.047)$  & $0.042\ (0.035;0.047)$& $0.042\ (0.036;0.048)$& $0.042\ (0.037;0.047)$ \\
\hline 
\end{tabular}
\end{center}
\end{table}

\begin{figure}[t]
\begin{center}
\includegraphics[scale=0.6]{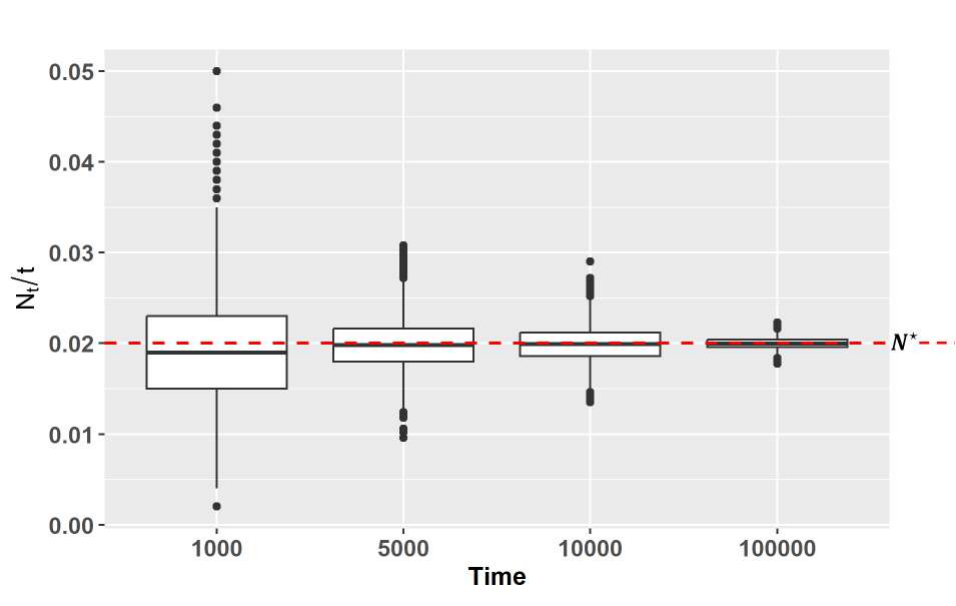}
\caption{\textbf{Consistency of $\frac{N_H}{H}$ as an estimator of $N^\star$ for $10000$ samples and $H=1000$, $H=5000$, $H=10000$, $H=100000$. True value is $N^\star=0.02$}}
\label{fig2}
\end{center}
\end{figure}

\paragraph{Limit moment of $Y$} To estimate $N^{\star}$, we use a different strategy as the total number of new isolated $Y_{(0,H]}$ is supposed to be observed. Therefore for each sampled trajectory we use the single estimate $\frac{Y_{(0,H]}}{H}=\frac{N_H}{H}$ to approximate the limit moment $N^\star$. 
The consistency of this estimation as $H$ increases is illustrated in Figure \ref{fig2}.

\begin{figure}[t]
\begin{center}
\includegraphics[scale=0.12]{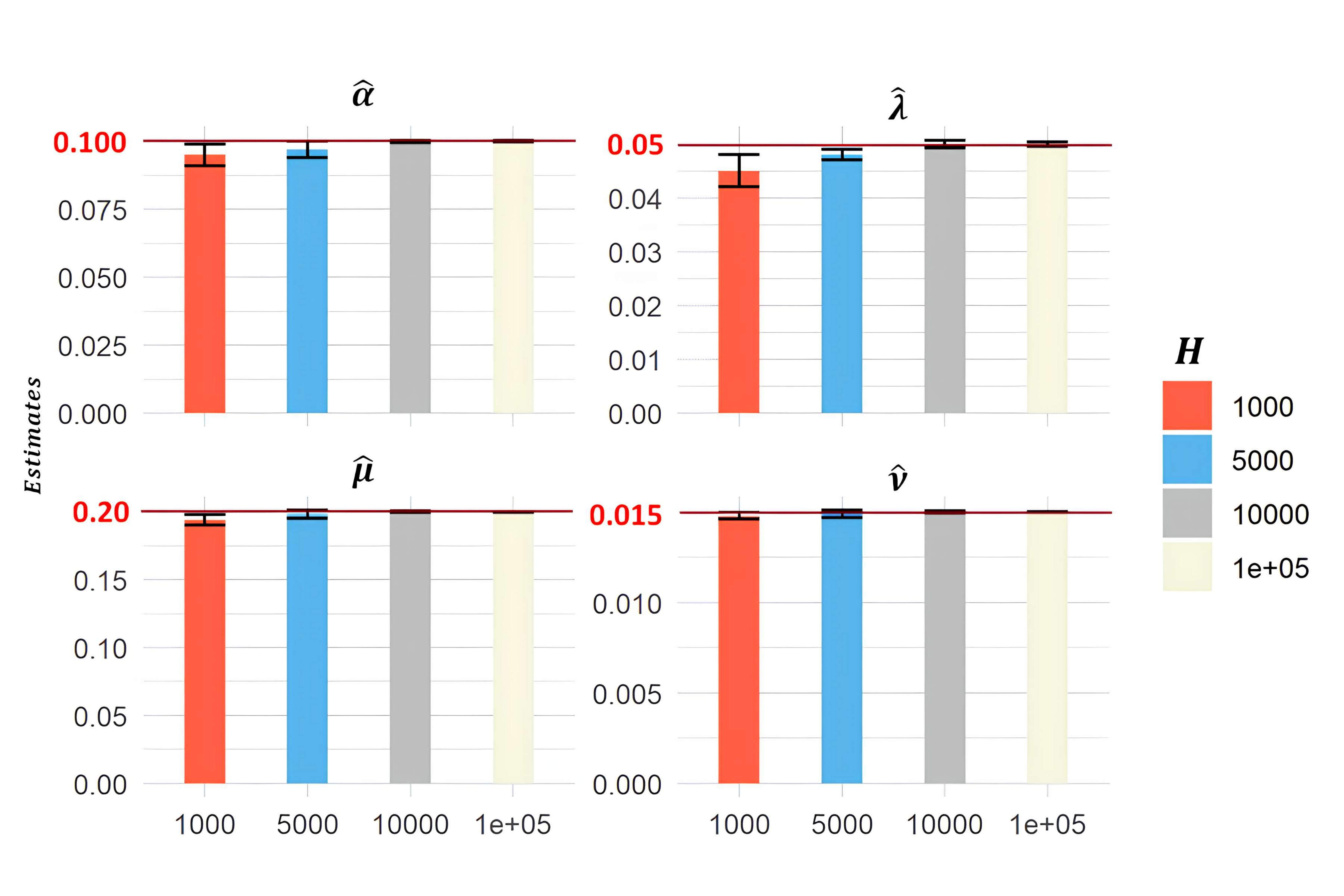}
\caption{\textbf{Parameter estimates and their $95\%$ empirical confidence intervals for $10000$ trajectories of the exposed-infected process and $H=1000$, $H=5000$, $H=10000$, $H=100000$. True values are $\lambda=0.05, \mu=0.2, \alpha=0.1$, and $\nu=0.015$.}}
\label{fig3}
\end{center}
\end{figure}

\paragraph{Estimation of the main parameters} Figure \ref{fig3} shows how the different errors described in the previous paragraph combine when estimating the parameters of interest $\alpha$, $\lambda$, $\mu$, $\nu$ from plugging $\overline{E_{H}}_{N_{MC}}$, $\overline{I_{H}}_{N_{MC}}$, $\overline{EI_{H}}_{N_{MC}}$ and $\frac{Y_{(0,H]}}{H}$ into to formulas given in Theorem \ref{th3} (numerical values displayed in Table~\ref{tableH::1} in Appendix \ref{sec:appx-num}).  The confidence intervals of the parameters are calculated taking into account only the variation of $N^\star$ (from $10000$ values of $N^\star$ and $ E^*,  I^*,  R^*$ are fixed at $\overline{E_{H}}_{N_{MC}}$, $\overline{I_{H}}_{N_{MC}}$, $\overline{EI_{H}}_{N_{MC}}$ respectively  for each H.
As expected, the performance significantly improves as $H$ increases, hence as the number of available observations increases. Observations on a time window of $10000$ days yields very good performance.
%
\subsection{Impact of the truncation of the state space}
\label{sec:trunc}
%
In order to run the Baum-Welch algorithm to estimate the transition probabilities $(p_{(e,i),(e',i')}, (e,i),(e',i') \in \mathbb{N}^2 )$ of the discrete time Markov chain $(I_n, E_n)$, truncation of the infinite state space and transition matrix is necessary. We are interested in low prevalence diseases, where counts of reported daily new isolated cases remain low. For some truncation value $N$, the corresponding truncated transition matrix $(p_{(e,i),(e',i')}^{(N)}, 0 \leq e,i, e',i' \leq N$ is defined by the true values for all $0\leq e,i, e',i'\leq N$ such that $(e',i')\neq (N,N)$, and a correction  for the $p_{(e,i),(N,N)}^{(N)}$ to obtain transition matrices for $i=N$, namely
$$p_{(e,i),(N,N)}^{(N)} = 1- \sum_{e'=0}^N \sum_{i'=0}^{N-1} p_{(e,i),(e',i')}^{(N)},$$
for $0\leq e,i\leq N$.

To investigate the impact of state truncation, we first simulated $10000$ trajectories of $(E_t, I_t)_{ 0 \leq t \leq H}$ with parameters $\lambda= 0.05, \mu=0.2, \alpha= 0.1$, and $\nu=0.015$ over a temporal horizon $H=10000$ to estimate the truncated transitions $(p_{(e,i),(e',i'))}^{(N)}$  for  $(I_n, E_n)$ for time step $\Delta t=1$. Then we used these estimated transition probabilities to simulate 
$10000$ samples 
$(E_H^{(N)},I_H^{(N)})$ of the truncated chain for a long time horizon $H=100000$, estimated the limit moments from these samples as in Section \ref{sec:complete}. Results are displayed in Table \ref{table2} and show that estimations can be very accurate with as low truncation thresholds as $N=5$. This is mainly due to our choice of parameters consistent with a low incidence disease.
\begin{table}[t]
\begin{center}
\caption{Estimates of the limit moments $E^*$, $I^*$ and $R^*$ (with $95\%$ confidence interval)
with samples obtained from the truncated discrete-time transition matrices $(p_{(e,i),(e',i')}^{(N)}$ for different values of $N$ and horizon $H=100000$ days.
True values are $E^*=0.2$, $I^*=0.1$ and $R^*=0.042$.}
\label{table2}
\begin{tabular}{@{}lrrr@{}}
\hline
Estimator & $N=3$ & $N=4$ &  $N=5$ \\
\hline
$\overline{E_{H}^{(N)}}_{N_{MC}}$  & $0.2020 (0.1932; 0.2108)$  & $0.2016 (0.1926;0.2106)$& $0.2014 (0.1924;0.2104)$ \\
$\overline{I_{H}^{(N)}}_{N_{MC}}$  & $0.0985 (0.0922; 0.1048)$ & $0.1015(0.0951;0.1079)$ & $0.1010 (0.0945;0.1075)$\\
$\overline{EI_{H}^{(N)}}_{N_{MC}}$  & $0.0429 (0.0363;0.0495)$  & $0.0420(0.0360;0.0480)$& $0.0420 (0.0365;0.0475)$ \\
\hline 
\end{tabular}
\end{center}
\end{table}
%
\subsection{Estimation based on observations of cumulated new isolated cases}
\label{HMM}
%
We will now focus on the HMCMM framework, where only the new cumulative reported cases $(Y_n)$ are observed. 

\begin{figure}[t]
\begin{center}
\includegraphics[scale=0.6]{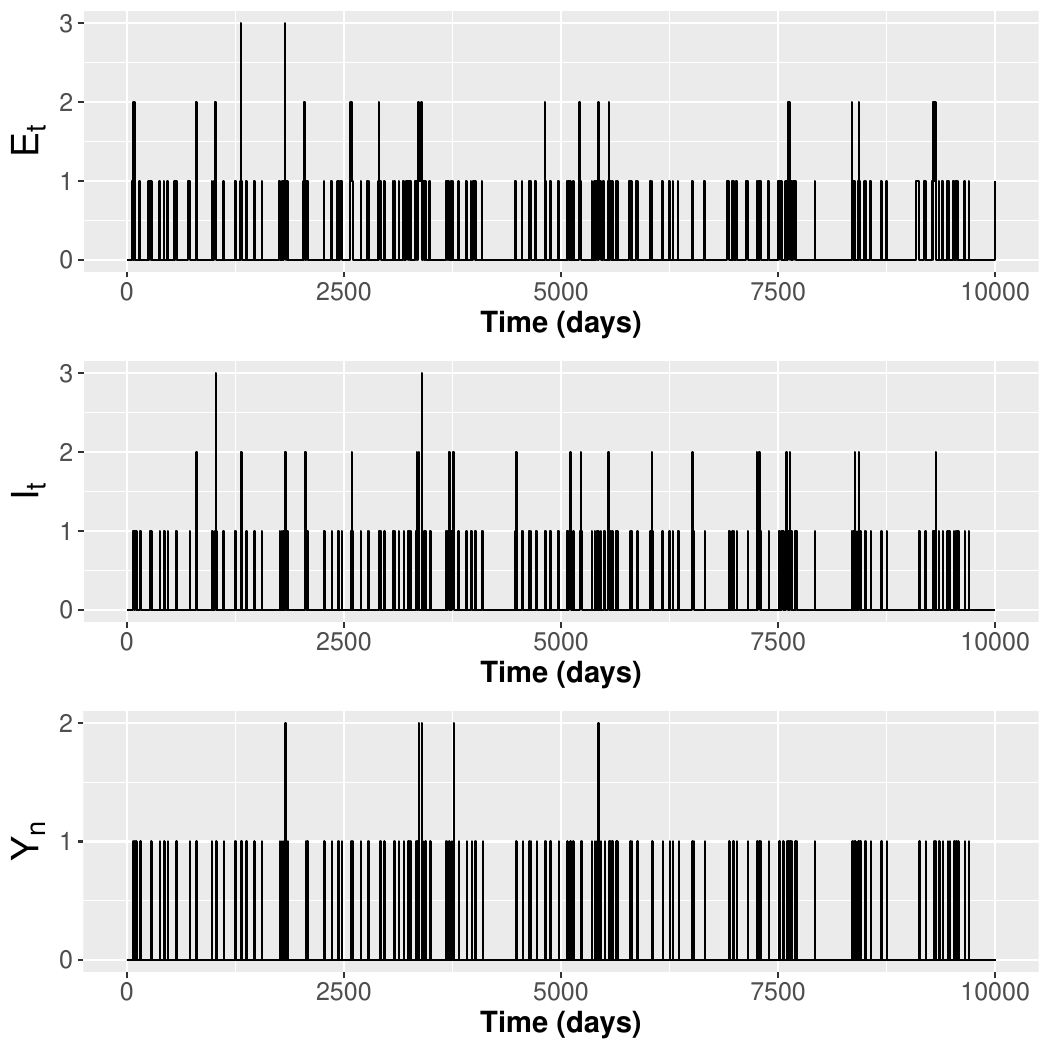}
\caption{\textbf{A sample trajectory of the exposed-infected process $(E_t,I_t)$ over a time horizon $H=10000$ days, with parameters ${\lambda=0.05, \mu=0.2, \alpha=0.1}$ and ${\nu=0.015}$ and the corresponding sample trajectory of the cumulative number of new isolated $Y_n$ over periods of $\Delta t = 1$ day.}}
\label{fig4}
\end{center}
\end{figure}

\paragraph{Observation samples} To do this, we first simulated $100$ sample trajectories of the true continuous-time process $(E_t,I_t, Y_{(0,t]})$ with parameters $\lambda=0.05, \mu=0.2, \alpha=0.1$ and ${\nu=0.015}$, over a time period $H$, see Figure \ref{fig4} for an example. Then we ran the adapted Baum-Welsh algorithm on the truncated state space with truncation parameter $N$ (for $E$ and $I$), and $M=\max\{Y_n\}$ (for the observations) for each observation sequence $(Y_n)$ in order to obtain the estimated truncated transition matrix. Finally, we ran the estimation procedure described in the previous parts to obtain estimations of $\alpha$, $\lambda$, $\mu$ and $\nu$ by sampling from the estimated truncated transition matrix.
\paragraph{Initialization of the Baum-Welch algorithm} The Baum-Welch algorithm is known to be sensitive to the choice of initial parameters. Therefore, it is important to choose initial parameters that are close to and in the same order of magnitude as the true parameter values as much as possible. This can be done by using prior knowledge of the system under study or by performing initial simulations to obtain reasonable values for the parameters. In our study, we decided to select the initial parameters in the following ranges: $ \lambda^{(0)}\in [0.04,0.07], \mu^{(0)}\in [0.185,0.25], \alpha^{(0)}\in [0.09,0.130] $ and $ \nu^{(0)}\in [0. 013,0. 02]$ and generate $15$ initial tuples of values within these ranges. 

For each tuple $(\lambda_i^{(0)},\mu_i^{(0)},\alpha_i^{(0)},\nu_i^{(0)})$ for $i=1,\ldots,15$, we run parallelized Monte Carlo simulations on $10000$ different trajectories to estimate the coefficients of the discrete time transition matrix $P^{(0)}$ and emission probabilities $\psi^{(0)}$. We then apply the space truncation approximation to these values
\begin{align*}
p_{(e,i)(N,N)}^{(0)}&= 1- \sum_{e'=0}^{N-1} \sum_{i'=0}^{N-1} p_{(e,i)(e',i')},\\
\psi_{(e,i,j)}(M)^{(0)}&= 1- \sum_{y=0}^{M-1} \psi_{(e,i,j)}(y),
\end{align*}
for $e,i,j = 0, \ldots,N$.
For the other parameters of our iterative scheme, we set $\Delta t = 1$,  the maximum number of iterations to $500$ and the stopping criterion  to $10^{-9}$. 

\paragraph{Parameter estimation} We run the adapted Baum-Welch algorithm for each value of $(\lambda_i^{(0)},\mu_i^{(0)},\alpha_i^{(0)},\nu_i^{(0)})$ for $i=1,\ldots,15$, and finally choose the transition matrix that maximises the likelihood of the observations.
$$\hat{p}^m= max_{\hat{p}^n} \mathbb{P}(Y|M^n). $$ 
We then use this optimal matrix $\hat{p}^m$ to simulate the Markov chain $(E_n,I_n)$ with a sufficiently large size of $n$ to recover estimates of the limit moments $E^*,I^*$ and $R^*$ and estimate $N^\star$ by $\frac{1}{n}\sum_{k=1}^nY_k$. Finally, we applied Theorem \ref{th3} to estimate the parameters of the exposed-infected model $\lambda, \alpha, \mu $ and $\nu$ from the estimated limit moments.

The impact of the choice of truncation parameters $N$ and time horizon $H$ is shown in Figures \ref{fig6} and \ref{fig7} (numerical values are given in Tables \ref{tableH::2} and \ref{tableH::3} in Appendix \ref{sec:appx-num}). 
\begin{figure}[t]
\begin{center}
\includegraphics[scale=0.12]{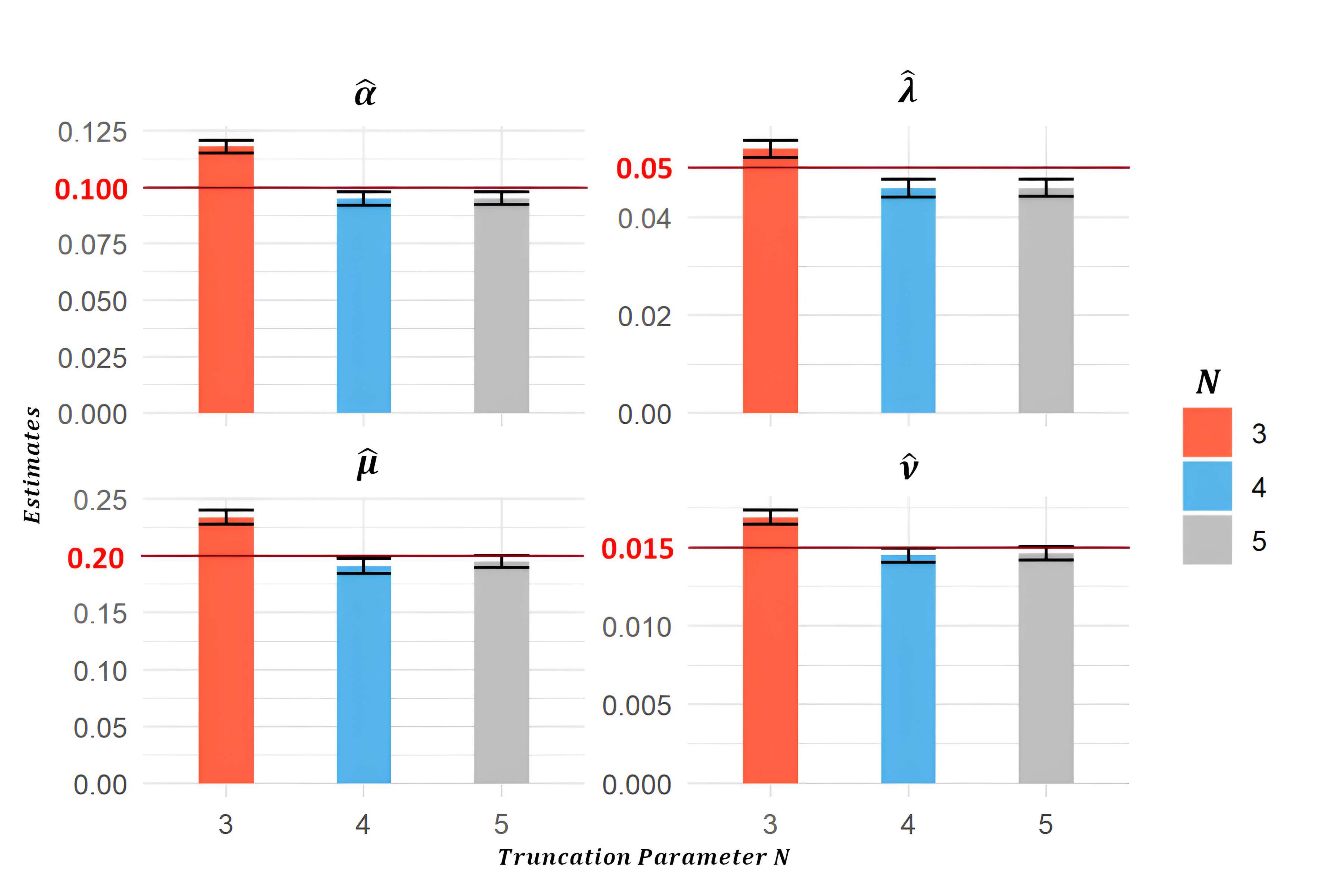}
\caption{\textbf{Estimation of parameters for different values of $N$ from observed new isolated on a trajectory with horizon $H=1000$  ($100$ replications, true values $\lambda=0.05, \mu=0.2, \alpha=0.1$ and $\nu=0.015$).}}
\label{fig6}
\end{center}
\end{figure}
\begin{figure}[t]
\begin{center}
\includegraphics[scale=0.12]{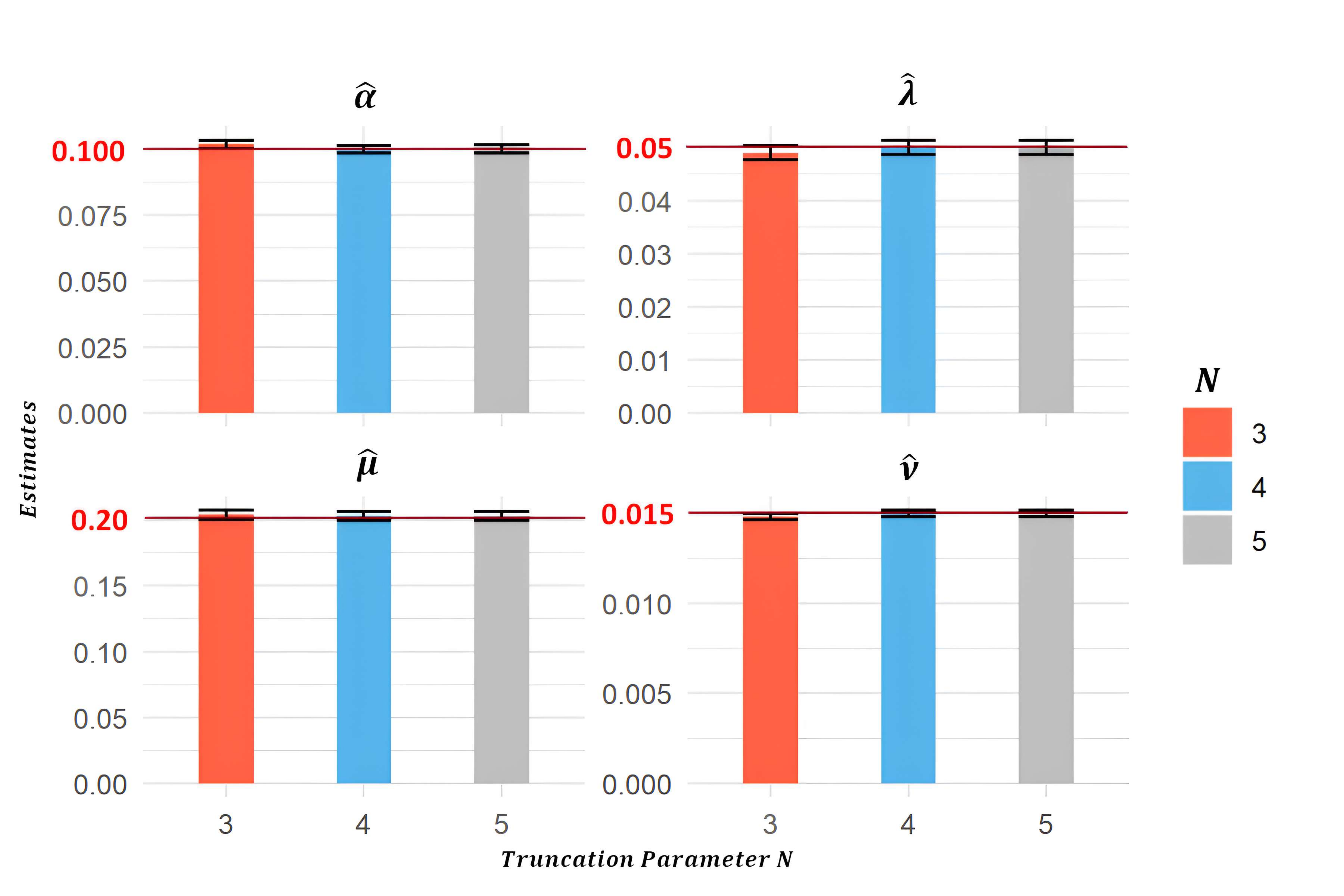}
\caption{\textbf{Estimation of parameters for different values of $N$ from observed new isolated on a trajectory with horizon $H=10000$  ($100$ replications, true values $\lambda=0.05, \mu=0.2, \alpha=0.1$ and ${\nu=0.015}$).}}
\label{fig7}
\end{center}
\end{figure}
On the one hand, we note that the parameter estimates exhibit minimal variation irrespective of the chosen truncation parameter $N$. Specifically, when setting $H=10000$, across the three tested values of $N$, the parameter estimates remain consistently stable and closely aligned. This suggests that the choice of state space truncation exerts negligible influence on the estimation process. Such stability is attributed to the predominant occurrence of states $0$ and $1$ within the observed data. However, a large value of $N$ also increases the computation time required to obtain the results. On the other hand, the selection of the number of observations $n$ or equivalently horizon $H$ has a major influence on the estimation. Estimation is quite poor for $H=1000$, but quite accurate for $H=10000$. 

Our numerical experiments have shown that the identifiability of the parameters is obtained with our estimation strategy even under a quite poor observation scheme, providing there are enough observations. 
%
 \subsection{Model selection}
 \label{sec:choice}
%
In this section, we conduct a comparative analysis of two estimation frameworks based on different models, in order to help data driven model choice. The first framework corresponds to the one-compartment model (model 1) studied in \cite{bouzalmat2024parameter}, characterized by three fundamental parameters $(\lambda, \mu,$ and $\nu$) and corresponding to a  Linear Birth and Death  process with Immigration (LBDI). In contrast, the second framework involves the two-compartment exposed-infected model (model 2) introduced in this paper, which introduces an additional parameter ($\alpha$) into the estimation process. 

\paragraph{Sampling from model 1}
Firstly, we generate $100$ trajectories the LBDI process, starting from $I_0=0$ infected, until a tome horizon $H=10000$. Then, for each trajectory we extract the observations that corresponded to the cumulative count of newly isolated cases $Y_{n}^1$, employing a fixed time-lapse $\Delta t = 1$. These observations were then employed to estimate the parameters of both models through the Hidden Markov Model (HMM) procedure introduced in \cite{bouzalmat2024parameter} for the LBDI process, and through the framework developed in this paper for the exposed-infected model. Global estimation performance is evaluated through the Bayesian Information Criterion (BIC). Results are shown on Table \ref{table4} for parameter values of $\lambda = 0.05, \mu = 0.5$, and $\nu = 0.01$ and on Table \ref{table6} for parameter values of $\lambda = 0.1, \mu = 0.2$, and $\nu = 0.015$.
\begin{table}[t]
\begin{center}
\caption{Parameter estimation results (with empirical standard errors for $100$ replications) for models 1 and 2 using observations generated by model 1, time horizon $H=10000$, truncation parameter $N=4$ and true parameters $\lambda = 0.05, \mu = 0.5$, and $\nu = 0.01$}
\label{table4}
\begin{tabular}{@{}lrr@{}}
\hline
Parameters & Model 1 & Model 2 \\
\hline
$\hat\lambda^n$   & $0.05 (4.13\cdot10^{-3})$  &$0.048 (3.12\cdot10^{-3})$ \\
$\hat\mu^n$   & $0.50 (6.37\cdot10^{-3}) $ &$0.52\phantom{0} (9.66\cdot10^{-3})$ \\
$\hat\nu^n$  &  $0.01 (1.24\cdot10^{-4})$  & $0.01\phantom{0}  (1.08\cdot10^{-4})$ \\
$\hat\alpha^n$   & $-$  & $2.09\phantom{0}  (4.20\cdot10^{-2})$  \\
BIC  & ${\bf 310.6054}$  & $ 401.5587$  \\
\hline 
\end{tabular}
\end{center}
\end{table}
\begin{table}[t]
\begin{center}
\caption{Parameter estimation results (with empirical standard errors for $100$ replications) for models 1 and 2 using observations generated by model 1, time horizon $H=10000$, truncation parameter $N=4$ and true parameters $\lambda = 0.1, \mu = 0.2$, and $\nu = 0.015$}
\label{table6}
\begin{tabular}{@{}lrr@{}}
\hline
Parameters & Model 1 & Model 2 \\
\hline
$\hat\lambda^n$   & $0.05\phantom{0} (3.70\cdot10^{-3})$  &$0.016 (5.65\cdot10^{-3})$ \\
$\hat\mu^n$   & $0.20\phantom{0} (1.86\cdot10^{-3}) $ &$0.33\phantom{0} (1.70\cdot10^{-2})$ \\
$\hat\nu^n$  &  $0.015 (1.50\cdot10^{-4})$  & $0.019 (3.56\cdot10^{-4})$ \\
$\hat\alpha^n$   & $-$  & $0.184 (8.24\cdot10^{-3})$  \\
BIC  & ${\bf 323.2059}$  & $ 375.3678$  \\
\hline 
\end{tabular}
\end{center}
\end{table}
For both parameter sets, the results clearly demonstrate that Model 1 has successfully and accurately estimated the parameters associated with its structure, providing compelling evidence of Model 1 identifiability in the context of our study. Furthermore, Model 1 exhibits a lower BIC value than model 2, leading to the correct model choice.

\paragraph{Sampling from model 2 with short incubation period}
Conversely, we conducted simulations under model 2, starting from the initial state $(0,0)$, with parameters $\alpha=2, \lambda = 0.05, \mu = 0.5$, and $\nu = 0.01$, over the identical time frame $H=10000$. Then, we derived the sequence of observations $Y^2_{n}$, which was subsequently employed for parameter estimation within the two estimation frameworks as above. The results are presented in Table \ref{table5}.
\begin{table}[t]
\begin{center}
\caption{Parameter estimation results (with empirical standard errors for $100$ replications) for models 1 and 2 using observations generated by model 2, time horizon $H=10000$, truncation parameter $N=4$ and true parameters $\alpha=2, \lambda = 0.05, \mu = 0.5$, and $\nu = 0.01$}
\label{table5}
\begin{tabular}{@{}lrr@{}}
\hline
Parameters & Model 1 & Model 2 \\
\hline
$\hat\lambda^n$   & $0.047 (3.04\cdot10^{-3})$  &$0.049 (2.80\cdot10^{-3}) $ \\
$\hat\mu^n$   & $0.48\phantom{0} (5.02\cdot10^{-3}) $ &$0.5\phantom{00} (9.23\cdot10^{-3})$ \\
$\hat\nu^n$  &  $0.01\phantom{0} (9.81\cdot10^{-5})$  & $0.01\phantom{0} (1.29\cdot10^{-4}) $ \\
$\hat\alpha^n$   & $-$  & $2\phantom{.000} (3.78\cdot10^{-2})$  \\
BIC  & ${\bf 242.7923}$  & $ 411.5036$  \\
\hline 
\end{tabular}
\end{center}
\end{table}
Although parameters are successfully estimated by the model 2 specific procedure, it has a higher BIC value than model 1 and is thus not preferred to Model 1. Note that model 1 still provides estimates close to the true values for $\lambda$, $\mu$ and $\nu$. This is because we chose a comparatively high value of $\alpha$, meaning a short incubation period making it harder to distinguish between the two models, especially with our comparatively poor observation scheme. Indeed, a high value of $\alpha$ implies a rapid transition of individuals from the exposed to the infected state and the two models become effectively equivalent, see \cite{champredon2018equivalence,jiao2020dynamics,kuznetsov1994bifurcation}.

\paragraph{Sampling from model 2 with long incubation period}
In our last scenario, we conducted simulations under model 2, starting from the initial state $(0,0)$, with parameters $\alpha=0.1, \lambda = 0.05, \mu = 0.2$, and $\nu = 0.015$, over the identical time frame $H=10000$. We then derived the sequence of observations $Y^2_{n}$, which was subsequently employed for parameter estimation within the two estimation frameworks as above. The results are presented in Table 
\ref{table7}.
\begin{table}[t]
\begin{center}
\caption{Parameter estimation results (with empirical standard errors for $100$ replications) for models 1 and 2 using observations generated by model 2, time horizon $H=10000$, truncation parameter $N=4$ and true parameters $\alpha=0.1, \lambda = 0.05, \mu = 0.2$, and $\nu = 0.015$}
\label{table7}
\begin{tabular}{@{}lrr@{}}
\hline
Parameters & Model 1 & Model 2 \\
\hline
$\hat\lambda^n$   & $0.010 (1.46\cdot10^{-3})$  &$0.050 (1.35\cdot10^{-3}) $ \\
$\hat\mu^n$   & $0.191 (1.59\cdot10^{-3}) $ &$0.205 (3.40\cdot10^{-3})$ \\
$\hat\nu^n$  &  $0.018 (1.70\cdot10^{-5})$  & $0.015 (1.79\cdot10^{-4}) $ \\
$\hat\alpha^n$   & $-$  & $0.100 (1.43\cdot10^{-3})$  \\
BIC  & $381.4615$  & ${\bf 314.0301}$  \\
\hline 
\end{tabular}
\end{center}
\end{table}
Again, parameters are successfully estimated by the model 2 specific procedure. However, this time the right model is selected, and the estimation of parameter $\lambda$ from model $1$ is quite poor, although estimates of $\mu$ and $\nu$ are still precise enough. This time, we chose a low value for $\alpha$, corresponding to a long incubation period and making both models more different. 
\section{Conclusion}\label{sec5}
We studied a two-compartment epidemic model (exposed-infected) to characterize the transmission dynamics of infectious diseases with low prevalence. We provided an estimation methodology for the parameters of the model when only counts of newly isolated individuals are observed. Our procedure involves parameter expressions in terms of limit moments of the process and the adaptation of the Baum-Welch algorithm within the framework of hidden Markov multi-chains. Our numerical simulations further demonstrated the effectiveness of our estimation methodology regarding identifiability, parameter estimation and model choice when compared to a single compartment model.

This procedure requires a high number of observations (typically daily counts over more than 27 years) to be sharp, therefore it is still an open challenge to accurately estimate the parameters for real data sets with shorter data sequences.
\appendix
\section{Proof of the main results} 
\subsection{Proof of Lemma \ref{lemma2.1}} 
\label{proof::Lemma1}
\begin{proof}
In order to prove Lemma \ref{lemma2.1}, we use the infinitesimal characterization of the distribution of the joint process. 
For any initial states $(e,i)$, we use the usual notation $\mathbb{P}_{(e,i)}(\cdot)=\mathbb{P}(\cdot|(E_0, I_0)=(e,i))$.
For $h > 0$, $e,i,y,e',i'\in\mathbb{N}$ with $i\leq i'+y$, the Markov property yields
\begin{align*} 
\lefteqn{p_{(e,i),(e',i',y)}(t+h)}\\
&= \mathbb{P}_{(e,i)} \left((E_{t+h},I_{t+h}) =(e',i'), Y_{]0,t+h]} = y\right) \\ 
&= \mathbb{P}_{(e,i)} \big((E_{t+h},I_{t+h}) =(e',i'), Y_{]h,t+h]} = y,(E_{h},I_{h}) =(e+1,i), Y_{]0,h]} = 0\big)   \\ 
&+ \mathbb{P}_{(e,i)} \big((E_{t+h},I_{t+h}) =(e',i'), Y_{]h,t+h]} = y,(E_{h},I_{h}) =(e-1,i+1), Y_{]0,h]} = 0\big)\\
&+\mathbb{P}_{(e,i)}  \big((E_{t+h},I_{t+h}) =(e',i'),Y_{]h,t+h]} = y-1,(E_{h},I_{h}) =(e,i-1),  Y_{]0,h]} = 1\big)\\
&+ \mathbb{P}_{(e,i)} \big((E_{t+h},I_{t+h}) =(e',i'), Y_{]h,t+h]} = y,(E_{h},I_{h}) =(e,i), Y_{]0,h]} = 0\big)+o(h)\\
&= \mathbb{P}_{(e,i)} \left( (E_{h},I_{h}) =(e+1,i), Y_{]0,h]} = 0\right) \mathbb{P}_{(e+1,i)}\big( (E_{t},I_{t}) =(e',i'), Y_{]0,t]} = y\big) \\
&+ \mathbb{P}_{(e,i)}\big( (E_{h},I_{h}) =(e-1,i+1), Y_{]0,h]} = 0)\big)\mathbb{P}_{(e-1,i+1)}\big( (E_{t},I_{t}) =(e',i'),  Y_{]0,t]} = y\big) \\
&+ \mathbb{P}_{(e,i)}\big( (E_{h},I_{h}) =(e,i-1), Y_{]0,h]} = 1\big) \mathbb{P}_{(e,i-1)}\big( (E_{t},I_{t}) =(e',i'),Y_{]0,t]} = y-1\big)\\
& + \mathbb{P}_{(e,i)}\big( (E_{h},I_{h}) =(e,i), Y_{]0,h]} = 0\big) \mathbb{P}_{(e,i)}\big( (E_{t},I_{t}) =(e',i'), Y_{]0,t]} = y\big) +o(h)\\
&= \mathbb{P}_{(e,i)} \left( (E_{h},I_{h}) =(e+1,i)\right) \mathbb{P}_{(e+1,i)}\big( (E_{t},I_{t}) =(e',i'), Y_{]0,t]} = y\big) \\
&+ \mathbb{P}_{(e,i)}\big( (E_{h},I_{h}) =(e-1,i+1))\big)\mathbb{P}_{(e-1,i+1)}\big( (E_{t},I_{t}) =(e',i'),  Y_{]0,t]} = y\big) \\
&+ \mathbb{P}_{(e,i)}\big( (E_{h},I_{h}) =(e,i-1)\big) \mathbb{P}_{(e,i-1)}\big( (E_{t},I_{t}) =(e',i'),Y_{]0,t]} = y-1\big)\\
& + \mathbb{P}_{(e,i)}\big( (E_{h},I_{h}) =(e,i)\big) \mathbb{P}_{(e,i)}\big( (E_{t},I_{t}) =(e',i'), Y_{]0,t]} = y\big) +o(h).
\end{align*}
Using the infinitesimal characterization of the process $(E_t,I_t)_{t\geq 0}$, one obtains
\begin{align*} 
\lefteqn{p_{(e,i),(e',i',y)}(t+h)}\\
&= (\lambda i + \nu )h  p_{(e+1,i),(e',i',y)}(t) + \alpha e  h p_{(e-1,i+1),(e',i',y)}(t)\\
&\quad  + \mu i h p_{(e,i-1),(e',i',o-1)}(t)+ (1-((\lambda  + \mu ) i + \alpha e + \nu)h )p_{(e,i),(e',i',y)}(t)+o(h).
\end{align*}
The desired result is therefore achieved by division by $h$ and letting $h$ tend to $0$.
\end{proof}
\subsection{Proof of Theorem \ref{th3}}
\label{proof::Th3.1}
\begin{proof}
The Markov process $(E_t,I_t,Y_{(0,t]})_{t \geq 0 }$ with values in $\tilde{S}=\mathbb{N}\times \mathbb{N}\times \mathbb{N}$ has transition rates matrix $\tilde{Q} $ given by
\begin{align*}
\tilde{Q}_{(e,i,y),(e+1,i,y)}&= \lambda i + \nu, \\
\tilde{Q}_{(e,i,y),(e-1,i+1,y)}&= \alpha e\mathbf{1}_{e>0}, \\
\tilde{Q}_{(e,i,y),(e,i-1,y+1)}&=\mu i\mathbf{1}_{i>0}, \\
\tilde{Q}_{(e,i,y),(e,i,y)}&= - (( \lambda + \mu ) i + \alpha e + \nu ).
\end{align*}
for all  $(e,i,n)\in\tilde{S}$.

For any nonnegative function $f$ on $\tilde{S}$ and for all initial state $(e_0,i_0,0) \in \tilde{S}$, we have:
$$ \mathbb{E}_{(e_0,i_0)} [ f(E_t,I_t,Y_{(0,t]})]=\sum_{ (e,i,y) \in \tilde{S}} p_{(e_0,i_0)(e,i,y)}(t) f(e,i,y),$$
with the notation of Lemma \ref{lemma2.1}.
Using again the forward Kolmogorov equation, we obtain
\begin{align*}
\lefteqn{\dfrac{d}{d t} \mathbb{E}_{(e_0,i_0)} [ f(E_t,I_t,Y_{(0,t]})]}\\
&= \sum_{ (e,i,y)  \in \tilde{S}} \left(  \sum_{ (e',i', y') \in \tilde{S}} p_{(e_0,i_0)(e',i' , y')}(t) \tilde{Q}_{(e',i' , y')(e,i , y)} \right) f(e,i , y) \\
&=\sum_{ (e',i',y') \in \tilde{S}} p_{(e_0,i_0)(e',i' , y')}(t) \Big( ( \lambda i' +\nu) \big(f(e'+1,i',y' )-f(e',i',y' ) \big)\\
&\quad + \alpha e' \big( f(e'-1,i'+1,y' ) -f(e',i',y' ) \big) \\ 
 &\quad + \mu i' \big( f(e',i'-1,y'+1 ) -f(e',i',y' ) \big) \Big)  \\ 
 &= \mathbb{E}_{(e_0,i_0)} \big[(\lambda I_t +\nu) \big( f(E_t+1,I_t,Y_{(0,t]})) - f(E_t,I_t,Y_{(0,t]})) \big)\\
 &\quad + \alpha E_t \big( f(E_t-1,I_t+1,Y_{(0,t]}) - f(E_t,I_t,Y_{(0,t]}) \big) \\ 
 &\quad + \mu I_t \big( f(E_t,I_t-1,Y_{(0,t]}+1)- f(E_t,I_t,Y_{(0,t]}) \big) \big].
\end{align*}
Considering functions $f(e,i,y)$, such as the coordinate functions $f(e,i,y)=e$, $f(e,i,y)=i$, $f(e,i,y)=y$ together with the product functions $f(e,i,y)=ei$, $f(e,i,y)=e^2$ and $f(e,i,y)=i^2$, we can establish the following system of differential equations for the moments of the process
\begin{equation*}\label{eqC.1}
\begin{cases}
\begin{aligned}
\dfrac{d}{dt} \mathbb{E}_{(e_0,i_0)} [ E_t]&= - \alpha \mathbb{E}_{(e_0,i_0)} [ E_t] + \lambda\mathbb{E}_{(e_0,i_0)} [ I_t] + \nu, \\ 
\dfrac{d}{dt}  \mathbb{E}_{(e_0,i_0)} [ I_t]&= \alpha \mathbb{E}_{(e_0,i_0)} [ E_t] - \mu\mathbb{E}_{(e_0,i_0)} [ I_t], \\ 
\dfrac{d}{dt}  \mathbb{E}_{(e_0,i_0)} [Y_{(0,t]}]&= \mu\mathbb{E}_{(e_0,i_0)} [ I_t], \\ 
\dfrac{d}{dt}  \mathbb{E}_{(e_0,i_0)} [ E^2_t]&=(2 \nu + \alpha) \mathbb{E}_{(e_0,i_0)} [ E_t] + \lambda \mathbb{E}_{(e_0,i_0)} [ I_t]  \\ 
& \ \ \ -2 \alpha \mathbb{E}_{(e_0,i_0)} [ E^2_t] + 2 \lambda \mathbb{E}_{(e_0,i_0)} [ E_t I_t] + \nu,\\
 \dfrac{d}{dt}  \mathbb{E}_{(e_0,i_0)} [ E_t I_t]&= - \alpha \mathbb{E}_{(e_0,i_0)} [ E_t] + \nu \mathbb{E}_{(e_0,i_0)} [ I_t] +\alpha \mathbb{E}_{(e_0,i_0)} [ E^2_t]  \\
        & \ \ \ -(\mu + \alpha) \mathbb{E}_{(e_0,i_0)} [ E_t I_t] + \lambda \mathbb{E}_{(e_0,i_0)} [ I^2_t], \\ 
 \dfrac{d}{dt}  \mathbb{E}_{(e_0,i_0)} [ I^2_t]&= \alpha \mathbb{E}_{(e_0,i_0)} [ E_t] + \mu \mathbb{E}_{(e_0,i_0)} [ I_t] + 2 \alpha \mathbb{E}_{(e_0,i_0)} [ E_t I_t]\\
 &  \ \ \  -2 \mu \mathbb{E}_{(e_0,i_0)} [ I^2_t]. 
\end{aligned}
\end{cases}
\end{equation*}
By solving the above system, we obtain the following expressions if $\lambda\neq \mu$
\begin{align}
\mathbb{E}_{(e_0,i_0)} [ E_t] &= c_1 e^{ - \frac{\mu + \alpha + \sqrt{(\mu - \alpha)^2 + 4 \alpha \lambda}}{2} t}+ c_2 e^{\frac{-\mu - \alpha + \sqrt{(\mu - \alpha)^2 + 4 \alpha \lambda}}{2} t} + \dfrac{\mu \nu}{\alpha \left( \mu - \lambda \right)}, \nonumber\\ 
\mathbb{E}_{(e_0,i_0)} [ I_t] &= c_3 e^{ - \frac{\mu + \alpha + \sqrt{(\mu - \alpha)^2 + 4 \alpha \lambda}}{2} t}+ c_4 e^{\frac{-\mu - \alpha + \sqrt{(\mu - \alpha)^2 + 4 \alpha \lambda}}{2} t} + \dfrac{\nu}{ \mu - \lambda }\nonumber, \\
\mathbb{E}_{(e_0,i_0)} [Y_{(0,t]}] &=  c_5 + c_6 e^{ - \frac{\mu + \alpha + \sqrt{(\mu - \alpha)^2 + 4 \alpha \lambda}}{2} t}+ c_7 e^{\frac{-\mu - \alpha + \sqrt{(\mu - \alpha)^2 + 4 \alpha \lambda}}{2} t} + \dfrac{\mu \nu}{ \mu - \lambda } t, \nonumber\\ 
 \mathbb{E}_{(e_0,i_0)} [ E_t I_t] &= c_8 e^{-(\mu + \alpha) t} + c_9 e^{ - \frac{\mu + \alpha + \sqrt{(\mu - \alpha)^2 + 4 \alpha \lambda}}{2} t}+ c_{10} e^{\frac{-\mu - \alpha + \sqrt{(\mu - \alpha)^2 + 4 \alpha \lambda}}{2} t} \nonumber\\ 
 & + c_{11} e^{-(\mu + \alpha + \sqrt{(\mu - \alpha)^2 + 4 \alpha \lambda})t}+c_{12} e^{(-\mu - \alpha + \sqrt{(\mu - \alpha)^2 + 4 \alpha \lambda})t} \nonumber \\ 
  & + \dfrac{\mu \nu \left( \left(\mu +\alpha \right) \nu + \alpha\lambda \right)}{\alpha\left(\mu - \lambda \right)^ 2 \left(\mu+ \alpha\right)},\label{eqC.2}
\end{align}
where $c_i$ for $i=1,\ldots,12$ are real constants depending on the initial state.
Thus, if $\lambda < \mu $, then one has
\begin{align*}
 E^* := \lim_{t \rightarrow +\infty} \mathbb{E}_{(e_0,i_0)} [ E_t] &= \dfrac{\mu \nu}{\alpha \left( \mu - \lambda \right)}, \\
    I^* := \lim_{t \rightarrow +\infty} \mathbb{E}_{(e_0,i_0)} [ I_t] &= \dfrac{\ \nu}{ \mu - \lambda }, \\
     N^{\star} := \lim_{t \rightarrow +\infty} \dfrac{\mathbb{E}_{(e_0,i_0)} [ Y_{(0,t]}]}{t} &=\dfrac{\mu \nu}{ \mu - \lambda },\\ 
     R^* := \lim_{t \rightarrow +\infty} \mathbb{E}_{(e_0,i_0)} [ E_t I_t] &= \dfrac{\mu \nu \left( \left(\mu +\alpha \right) \nu + \alpha\lambda \right)}{\alpha\left(\mu - \lambda \right)^ 2 \left(\mu+ \alpha\right)},
\end{align*}
and these four equations form an invertible system as a function of the parameters $(\lambda, \mu, \alpha, \nu)$ which yields the expected result.
\end{proof}
%
\subsection{Proof of Theorem \ref{th4}} \label{proof::th3.2}
%
\begin{proof}
Denote by $\mathcal{X}= \mathbb{N}^3$ the state space of the chain $(X_n)_{n\in\mathbb{N}^*}$ and $x=(e,i,j)\in \mathcal{X}$. The aim is to maximise the likelihood of observations $ \mathbb{P}(Y_{1:T}= y_{1:T}| M^{n+1} )$. By applying the expectation maximisation (EM) algorithm to the maximisation of this probability, the following auxiliary function is maximized   
\begin{align*}
\lefteqn{ \Gamma_{Y_{1:T}}(M^{n},M^{n+1})}\\
 &= \sum_{x_{1:T}  \in  \mathcal{X}^T} log \mathbb{P}(Y_{1:T}= y_{1:T}, X_{1:T}= x_{1:T}| M^{n+1})\mathbb{P}(X_{1:T}= x_{1:T}|Y_{1:T}, M^{n}). 
\end{align*}
Knowing that
$$\mathbb{P}(Y_{1:T}= y_{1:T}, X_{1:T}= x_{1:T}| M^{n+1})=\rho_{x_{1}}^{n+1} \left( \prod_{t=1}^{T-1} Q_{x_{t},x_{t+1}}^{n+1} \right) \left(\prod_{t=1}^{T} \psi_{x_{t}}^{n+1}(y_t) \right), $$
the function $\Gamma_{Y_{1:T}} $ is rewritten as follows 
\begin{align*}
\lefteqn{\Gamma_{Y_{1:T}}(M^{n},M^{n+1})}\\
 &= \sum_{x_{1:T}  \in  \mathcal{X}^T} log \rho_{x_{1}}^{n+1} \mathbb{P}(X_{1:T}= x_{1:T}|Y_{1:T}, M^{n}) \\
& + \sum_{x_{1:T}  \in  \mathcal{X}^T} \left( \sum_{t=1}^{T-1} log Q_{x_{t},x_{t+1}}^{n+1} \right)\mathbb{P}(X_{1:T}= x_{1:T}|Y_{1:T}, M^{n})\\ 
& +  \sum_{x_{1:T}  \in  \mathcal{X}^T} \left(\sum_{t=1}^{T} log \psi_{x_{t}}^{n+1}(y_t) \right)\mathbb{P}(X_{1:T}= x_{1:T}|Y_{1:T}, M^{n})\\ 
&= \Gamma_{Y_{1:T}}^{\rho}(M^{n},M^{n+1})+\Gamma_{Y_{1:T}}^{Q}(M^{n},M^{n+1})+\Gamma_{Y_{1:T}}^{\psi}(M^{n},M^{n+1}).
\end{align*}
We can then see that the function $ \Gamma_{Y_{1:T}} $ can be decomposed into three functions with distinct parameters. Consequently, we can optimise these three functions analytically using Lagrange multipliers. Let $\mathcal{L}(\rho,\theta), \mathcal{L}(Q,\theta) $ and $\mathcal{L}(\psi,\theta)$ be the Lagrangians associated respectively with $\Gamma_{Y_{1: T}}^{\rho},\Gamma_{Y_{1:T}}^{Q}$ and $\Gamma_{Y_{1:T}}^{\psi}$. One has
\begin{align*}
\mathcal{L}(\rho,\theta)&= \sum_{(e,i,j)} log \rho_{(e,i,j)}^{n+1} \mathbb{P}(X_{1}= (e,i,j)|Y_{1:T}, M^{n}) + \theta \left( \sum_{(e',i',j')} \rho_{(e',i',j')}^{n+1}-1 \right),\\
 \mathcal{L}(Q,\theta) &=    \sum_{(e,i,j)}  \sum_{(e',i',j')}  \left( \sum_{t=1}^{T-1} log Q_{(e,i,j),(e',i',j')}^{n+1} \right)  \\ 
& \quad  \times  \mathbb{P}(X_{t}=(e,i,j), X_{t+1}=(e',i',j') |Y_{1:T}, M^{n}) \\ 
& \quad + \sum_{(e,i,j)} \theta_{(e,i,j)} \left( \sum_{(e',i',j')} Q_{(e,i,j),(e',i',j')}^{n+1}-1 \right),\\
\mathcal{L}(\psi,\theta) &=    \sum_{(e,i,j)}  \left( \sum_{t=1}^{T} log \psi_{(e,i,j)}^{n+1}(y) \right)\mathbb{P}(X_{t}=(e,i,j) |Y_{1:T}, M^{n})\mathbf{1}_{y_t = y} \\
& \quad + \sum_{(e,i,j)} \theta_{(e,i,j)} \left( \sum_{o} \psi_{(e,i,j)}^{n+1}(o)  -1 \right). 
\end{align*}
By setting the first derivatives of the Lagrangians to zero,  we can show that $ \Gamma_{Y_{1:T}} $ is maximel when
\begin{align*}
Q_{(e,i,j),(e',i',j')}^{n+1} &= \dfrac{\sum_{t=1}^{T-1}\xi^n_{(e,i,j),(e',i',j')}(t)}{\sum_{t=1}^{T} \gamma^n_{(e,i,j)}(t) } \delta_{i'=j}, \\ 
\psi_{(e,i,j)}^{n+1}(y) &= \dfrac{\sum_{t=1}^{T} \mathbf{1}_{y_t = y} \gamma^n_{(e,i,j)}(t)}{\sum_{t=1}^{T} \gamma^n_{(e,i,j)}(t)},\\
\rho_{e,i,j}^{n+1}&= \gamma^n_{(e,i,j)}(1).
\end{align*}
On the other hand, using Lemma {\ref{lemme2}}, we can easily deduce that
$$ p_{(e,i),(e',i')}^{n+1}= \dfrac{\sum_{j'\in\mathbb{N} }Q^{n+1}_{(e,i,j),(e',j,j')}\rho_{e,i,j}^{n+1} }{\sum_{j\in\mathbb{N}}\rho_{e,i,j}^{n+1}},$$
hence the result.
\end{proof}
%
\section{Additional numerical results}\label{sec:appx-num}
In this section give some additional simulation result tables, corresponding to the figures displayed in Section \ref{sec4}.
\begin{table}[t]
\begin{center}
\caption{Parameter estimates and their $95\%$ empirical confidence intervals for $10000$ trajectories of the exposed-infected process and $H=1000$, $H=5000$, $H=10000$, $H=100000$. True values are $\lambda=0.05, \mu=0.2, \alpha=0.1$, and $\nu=0.015$.}
\label{tableH::1}
\begin{tabular}{@{}lrc@{}}
\hline
&\multicolumn{2}{c}{Time horizon $H=1000$}\\
\cline{2-3}
Parameter & Estimation & Confidence interval (95\%) \\
\hline
$\lambda$  & $0.045$  & $[0.042,0.048]$ \\
$\mu$  & $0.194 $ & $[0.190,0.198]$ \\
$\alpha$  & $0.095$  & $[0.091,0.099]$ \\
$\nu$ & $0.0148$  & $[0.0146,0.0150]$  \\
\hline
&\multicolumn{2}{c}{Time horizon $H=5000$}\\
\cline{2-3}
Parameter & Estimation & Confidence interval (95\%) \\
$\lambda$  & $0.048$  & $[0.047,0.049]$ \\
$\mu$  & $0.198 $ & $[0.195,0.201]$ \\
$\alpha$  & $0.097$  & $[0.094,0.100]$ \\
$\nu$ & $0.0149$  & $[0.0147,0.0151]$  \\
\hline
&\multicolumn{2}{c}{Time horizon $H=10000$}\\
\cline{2-3}
Parameter & Estimation & Confidence interval (95\%) \\
$\lambda$   &$0.050$  & $ [0.0493,0.0507] $ \\
$\mu$  & $0.20$ & $ [0.1994,0.2006] $\\
$\alpha$  & $0.100$  & $ [0.0996,0.1004] $ \\
$\nu$  & $0.015$  & $ [0.01493,0.01507] $ \\
\hline 
&\multicolumn{2}{c}{Time horizon $H=100000$}\\
\cline{2-3}
Parameter & Estimation & Confidence interval (95\%) \\
\hline
$\lambda$  & $0.050$  & $[0.0496,0.0504]$ \\
$\mu$  & $0.20 $ & $[0.1998,0.2002]$ \\
$\alpha$  & $0.10$  & $[0.0999,0.1001]$ \\
$\nu$ & $0.015$  & $[0.01497,0.01503]$  \\
\hline
\end{tabular}
\end{center}
\end{table}
\begin{table}[t]
\begin{center}
\caption{HMM-based estimation of parameters (with empirical standard errors) for different values of the truncation parameter $N$ on a trajectory with horizon $H=1000$ for $100$ replications.}
\label{tableH::2}
\begin{tabular}{@{}lrrr@{}}
\hline
Parameter & $N=3$ & $N=4$ & $N=5$\\
\cline{2-4}
$\hat\lambda^n$   & $0.054\phantom{0}(1.75\cdot10^{-3})$  &$0.046\phantom{0}(1.82\cdot10^{-3}) $ & $0.046\phantom{0}(1\cdot78\cdot10^{-3})$\\
$\hat\mu^n$   & $0.234\phantom{0}(6.10\cdot10^{-3}) $ &$0.191\phantom{0}(6.53\cdot10^{-3})$ &$0.195\phantom{0}(5.36\cdot10^{-3})$\\
$\hat\alpha^n$   & $0.118\phantom{0}(2.84\cdot10^{-3})$  & $0.095\phantom{0}(2.99\cdot10^{-3})$ &$0.095\phantom{0}(2.78\cdot10^{-3})$ \\
$\hat\nu^n$  & $0.0169 (4.46\cdot10^{-4})$  & $0.0145 (4.59\cdot10^{-4})$  &$0.0146 (4.30\cdot10^{-4})$\\
\hline 
\end{tabular}
\end{center}
\end{table}
\begin{table}[t]
\begin{center}
\caption{HMM-based estimation of parameters (with empirical standard errors) for different values of the truncation parameter $N$ on a trajectory with horizon $H=10000$ for $100$ replications.}
\label{tableH::3}
\begin{tabular}{@{}lrrr@{}}
\hline
Number of observations& $n=10000 $  \\
\hline
Parameter & $N=3$ & $N=4$ & $N=5$\\
\cline{2-4}
$\hat\lambda^n$   & $0.049\phantom{0}(1.35\cdot10^{-3})$  &$0.050 \phantom{0}(1.35\cdot10^{-3})$ & $0.050\phantom{0} (1.37\cdot10^{-3})$\\
$\hat\mu^n$   & $0.204\phantom{0}(3.34\cdot10^{-3}) $ &$0.205 \phantom{0}(3.40\cdot10^{-3})$ &$0.205\phantom{0} (3.11\cdot10^{-3})$\\
$\hat\alpha^n$   & $0.102\phantom{0}(1.40\cdot10^{-3})$  & $0.100 \phantom{0}(1.43\cdot10^{-3})$ &$0.100\phantom{0} (1.52\cdot10^{-3})$ \\
$\hat\nu^n$  &  $0.0148 (1.66\cdot10^{-4})$  & $0.015\phantom{0} (1.79\cdot10^{-4})$  &$0.015\phantom{0} (1.82\cdot10^{-4})$\\
\hline 
\end{tabular}
\end{center}
\end{table}
\newpage
\bibliographystyle{acm}
\bibliography{mybib}

\end{document}